\documentclass[12pt,preprint]{aastex}

\def\Gej{\Gamma_{\rm ej}}

\def\Lej{L_{\rm ej}}

\def\rhoejRS{\rho_{\rm ej}{\rm (RS)}}
\def\nejRS{n_{\rm ej}{\rm (RS)}}
\def\GejRS{\Gamma_{\rm ej}{\rm (RS)}}

\def\tobs{t_{\rm obs}}

\def\gm{\gamma_m}
\def\gc{\gamma_c}

\def\be{\begin{equation}}
\def\ee{\end{equation}}
\def\beq{\begin{eqnarray}}
\def\eeq{\end{eqnarray}}

\begin{document}

\title{Dynamics and Afterglow Light Curves of GRB Blast Waves \\
Encountering a Density Bump or Void} 

\author{Z. Lucas Uhm\altaffilmark{1}, Bing Zhang\altaffilmark{2,1,3}}

\altaffiltext{1}{Kavli Institute for Astronomy and Astrophysics, Peking University, 
Beijing 100871, China; uhm@pku.edu.cn, zhang@physics.unlv.edu}
\altaffiltext{2}{Department of Astronomy, School of Physics, Peking University, Beijing 100871, China}
\altaffiltext{3}{Department of Physics and Astronomy, University of Nevada, Las Vegas, NV 89154, USA}

\begin{abstract}
We investigate the dynamics and afterglow light curves of gamma-ray burst (GRB) blast waves 
that encounter various density structures (such as bumps, voids, or steps) in the surrounding 
ambient medium. We present and explain the characteristic response features that each type of 
density structures in the medium leaves on the forward shock (FS) and reverse shock (RS) dynamics, 
for blast waves with either a long-lived or short-lived RS. 
We show that, when the ambient medium density drops, the blast waves exhibit in some cases 
a period of an actual acceleration (even during their deceleration stage), due to adiabatic 
cooling of blast waves. Comparing numerical examples that have different shapes of bumps or voids, 
we propose a number of consistency tests that correct modeling of blast waves needs to satisfy. 
Our model results successfully pass these tests. Employing a Lagrangian description of blast waves, 
we perform a sophisticated calculation of afterglow emission. We show that, as a response to 
density structures in the ambient medium, the RS light curves produce more significant variations 
than the FS light curves. Some observed features (such as re-brightenings, dips, or slow wiggles) 
can be more easily explained within the RS model. We also discuss on the origin of these different 
features imprinted on the FS and RS light curves. 
\end{abstract}


%
%

\section{Introduction} \label{section:introduction}

Gamma-ray bursts (GRBs) with high quality afterglow data sometimes exhibit features
\citep[such as bumps and wiggles, e.g., GRB 021004, 030329,][]{holland03,lipkin04}
that deviate from the simple power law decay expected 
in the standard afterglow model \citep{meszarosrees97,sari98,chevalier00,gao13d}.
One of the most popular and appealing scenarios to account for these features is to 
invoke a sudden density change in the circumburst medium, e.g., a density bump or void. 
Indeed, a density bump has been invoked to interpret some re-brightening features observed
in some afterglow light curves \citep[e.g.][]{dailu02b,daiwu03}, and density fluctuation
has been invoked to interpret wiggles in some light curves \citep[e.g.][]{lazzati02}. 
However, more detailed numerical investigations have shown that the forward shock (FS) wave 
encountering a bumpy structure does not produce a prominent re-brightening signature 
in the light curves \citep[][]{nakargranot07,uhm07,gat13}.

In addition to the FS, a reverse shock (RS) is expected to propagate through the burst ejecta. 
The RS can be long-lived if there is a stratification in the ejecta's Lorentz factor $\Gej$ 
\citep[e.g.][]{rees98,sarimeszaros00,uhm07,genet07,uhm12}, or if the
central engine is long-lived \citep[e.g.][]{zhangmeszaros01a,dai04}.
An arising question here is then what should happen to the RS if it 
is present when the blast wave encounters an ambient density structure. 
This problem was first 
studied in \cite{uhm07} for the density bump case, briefly with a numerical example. 
\cite{uhm07} showed that when a blast wave with a long-lived RS meets bumps, 
the RS light curves exhibit prominent signatures while the FS light curves show 
no significant feature as mentioned above.

In this paper, we investigate this problem in detail\footnote{We adopt the scenario 
invoking a $\Gej$-stratification long-lived RS in this paper. The investigation may
also be generalized to the long-lived RS scenario invoking a long-lasting central engine.}.
We consider various shapes of density bumps in the ambient medium and extend the study 
to also include different shapes of density voids and step-like density structures.

As a blast wave propagates, the volume of shocked gas in the blast region increases. 
This implies that there is adiabatic conversion of the internal (thermal) energy of 
shocked gas into the kinetic bulk motion of the blast through $p dV$ work on itself, 
which we call ``adiabatic cooling'' of the blast wave. We prepare our numerical examples 
in such a way that comparisons among those can help us to demonstrate the role of adiabatic 
cooling and to propose a number of ``consistency tests'' that correct modeling of blast 
waves with a long/short-lived RS needs to satisfy.

%
%

\section{Dynamics and Afterglow Light Curves}

In order to find the blast wave dynamics of those problems described in Section~
\ref{section:introduction} with the density structures in the ambient medium, 
we make use of the semi-analytic formulation of relativistic blast wave, 
presented in \cite{uhm11} (hereafter U11). Instead of using a pressure balance across 
the blast wave, U11 applies the conservation laws of energy-momentum tensor and 
mass flux to the blast region between the FS and the RS, so that the two shock waves 
are ``connected'' through the blast via the relevant physics laws. The simple prescription 
of pressure balance does not provide an accurate solution for the blast wave dynamics, 
as shown in U11. U11 solves a continuity equation for the evolving ejecta flow 
to deal with its spherical expansion and radial spread-out and then determines 
which ejecta shell gets shocked by the RS at a certain radius. U11 also takes 
into account a variable adiabatic index $\kappa$ for shocked gas in the FS and RS regions. 
This is particularly important in describing the RS shocked gas since the RS strength 
evolves from the non-relativistic regime to mildly relativistic or even relativistic 
regime as the blast wave propagates (in a constant density medium), or vice versa (in 
a wind medium).

The formulation presented in U11 is applicable to a general class of GRB problems 
with a long-lived RS or with a short-lived RS, admitting an arbitrary 
radial stratification of the ejecta flow and the ambient medium. For this general 
kind of GRB blast waves, for which the FS dynamics should deviate from the self-similar 
solution of \cite{blandford76}, U11 provides an accurate numerical solution 
for both the FS and RS dynamics. This formulation also enables us to find a correct 
solution of the blast wave continuously from the initial coasting phase through 
the deceleration stage.

On the side of afterglow calculation, we make use of the Lagrangian description of 
the blast wave, described in \cite{uhm12} (hereafter U12). This description views 
the blast as being made of many different Lagrangian shells, piling up from the FS 
and RS fronts. Each shell in U12 has it own radius, pressure, energy density, 
adiabatic index, magnetic field, and electron energy distribution. This description 
is therefore significantly more sophisticated than the simple analytical afterglow 
model \citep{sari98} which assumes that the entire postshock region forms a single 
zone with the same radius, energy density and magnetic field, endowed with a single 
broken power-law distribution of electron energies.

As the blast wave propagates, U12 keeps track of the adiabatic evolution of each 
shell and finds its current energy density, adiabatic index, and magnetic field. 
As for the electron energy spectrum, a minimum Lorentz factor $\gm$ and a cooling 
Lorentz factor $\gc$ are numerically calculated for each shell at every calculation step 
by solving the full differential equation, Equation (17) in U12, which describes 
radiative and adiabatic cooling of electrons.\footnote{Note that $\gm$ in U12 was 
solved only for adiabatic loss that dominates over radiative loss for low energies. 
In this paper, we use the same full equation (including both cooling terms) to calculate 
both $\gm$ and $\gc$, for completeness.} 
U12 fully takes into account the curvature effect of every shell 
(the Doppler boosting of radial bulk motion and the spherical curvature of the shell), 
considering the fact that each shell has its own radius. 
The afterglow light curves are found by integrating over all Lagrangian shells, 
collecting photons emitted from each shell for every equal-arrival observer time $\tobs$. 
The importance of these detailed calculations is demonstrated, for instance, in our recent 
paper regarding on the shape of afterglow spectra where we show that a cooling break is 
very smooth and occurs gradually over several orders of magnitude in frequency and 
in observer time \citep{uhm14}.

%
%

\section{Numerical Examples} \label{section:numerical_examples}

We present 9 different numerical examples (named as 1a, 1b, 1c, 1d, 1e, 1f, 1g, 1h, 1i) with 
a long-lived RS and 5 different examples (named as 2a, 2c, 2d, 2f, 2g) with a short-lived RS. 
For all 14 examples, we keep the followings to be the same: (1) The burst is assumed to be 
located at a cosmological distance with redshift $z=1$. As in U12, 
we adopt a flat $\Lambda$CDM universe to calculate the luminosity distance, 
with the parameters $H_0=71$ km $\mbox{s}^{-1}$ $\mbox{Mpc}^{-1}$, 
$\Omega_{\rm m}=0.27$, and $\Omega_{\Lambda}=0.73$. 
(2) The microphysics parameters are $p=2.3$ (power law index of injected electrons), 
$\epsilon_e=10^{-1}$ (fraction of internal energy for accelerated electrons), and 
$\epsilon_B=10^{-2}$ (fraction of internal energy in magnetic fields) 
for the RS light curves, and $p=2.3$, $\epsilon_e=10^{-2}$, and 
$\epsilon_B=10^{-4}$ for the FS light curves.\footnote{Note that we have adopted different microphysics 
parameters for the FS and RS. The FS and RS regions originate from different sources, and strength 
of the two shocks are also significantly different. As in U12, we have chosen the emission parameters 
such that the FS and RS spectral fluxes become comparable to each other. By varying the parameters 
$\epsilon_B$ and/or $\epsilon_e$, we can further enhance or suppress the FS or the RS light curves. 
However, we fix the microphysics parameters for these 14 examples in order to focus our study on 
the response to density structures of the ambient medium.} 
(3) The ejecta has a constant kinetic luminosity $\Lej(\tau)=L_0=10^{53}~\mbox{erg}~\mbox{s}^{-1}$ 
for a duration of $\tau_b=10~\mbox{s}$, so that the total isotropic energy of the burst is 
$E_b=L_0\,\tau_b=10^{54}~\mbox{ergs}$.

\begin{figure}
\begin{center}
\includegraphics[width=10cm]{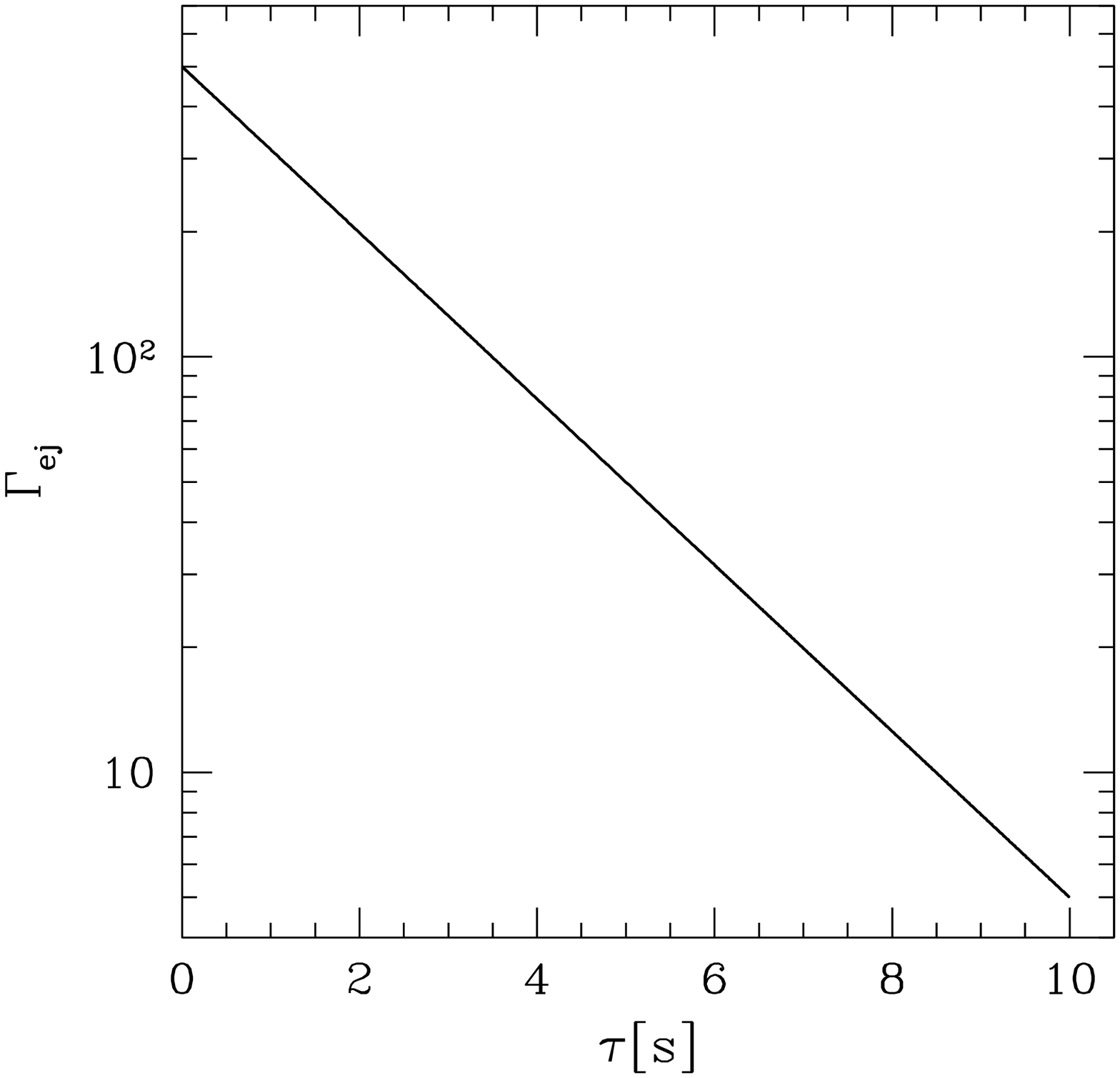}
\caption{
Stratification of the ejecta Lorentz factor $\Gej$ as a function of 
ejection time $\tau$ for numerical examples 1a, 1b, 1c, 1d, 1e, 
1f, 1g, 1h, and 1i. These 9 examples have a long-lived RS. 
} 
\label{fig:gej1}
\end{center}      
\end{figure}

For all 5 examples with a short-lived RS, the ejecta is assumed to emerge with a constant Lorentz factor 
$\Gej=300$. For all 9 examples with a long-lived RS, we assume a simple and monotonic stratification 
on the ejecta Lorentz factors $\Gej(\tau)=500 \times 10^{-\tau/5}$, which decreases exponentially 
from 500 to 5 for a duration of $\tau_b=10~\mbox{s}$; see Figure~\ref{fig:gej1}. 
A detailed investigation invoking various shapes of ejecta stratifications is presented in U12, 
where we explain how a density structure in the evolving ejecta affects the FS and RS dynamics 
and leaves its signature on the FS and RS light curves. For the present study, we keep one simple 
ejecta stratification and then examine what kind of imprints a density structure in the ambient medium 
leaves on the dynamics and afterglow light curves.

The difference among our examples goes on the radial profile of the ambient medium density. 
A constant density is assumed for the example 1a. The examples 1b, 1c, and 1d have a density bump 
in the ambient medium while the examples 1e, 1f, and 1g have a void in the medium. The example 1h 
has an increasing step in the density profile whereas the example 1i has a decreasing step in 
the profile. A detailed shape of these density structures is given below. 
The example names with the same alphabet but different numbers (1a vs. 2a, 1c vs. 2c, etc.) 
share the same density profile. In order to provide an efficient comparison among examples, 
we form groups of examples below.

\subsection{Group (1a/1b/1c)} \label{section:1a1b1c}

The example 1a has a constant density $\rho_1(r)/m_p=n_1(r)=1~\mbox{cm}^{-3}$ in the ambient medium. 
Here, $m_p$ is the proton mass. We denote this density by $n_0=1~\mbox{cm}^{-3}$ and use it as our 
fiducial base to place other density structures on. For comparison, we place all the structures at a same 
radius $r_0=3 \times 10^{17}~\mbox{cm}$. The example 1b has a density bump of Gaussian shape on top of 
the base, i.e., $n_1(r)=n_0+n_a\, e^{-(r-r_0)^2/(2\sigma^2)}$, where $n_a=4~\mbox{cm}^{-3}$ is the 
amplitude of the bump such that the peak density is 5 times the base density. As for the spatial 
spread of the bump, we use a 10 \% of its location; $\sigma=r_0/10$. This value will serve as our 
fiducial size $\sigma_0=3 \times 10^{16}~\mbox{cm}$. The example 1c has the same density profile 
as in the example 1b, except that the amplitude of the bump is $n_a=9~\mbox{cm}^{-3}$, so that 
the peak density is 10 times the base density.

\begin{figure}
\begin{center}
\includegraphics[width=17cm]{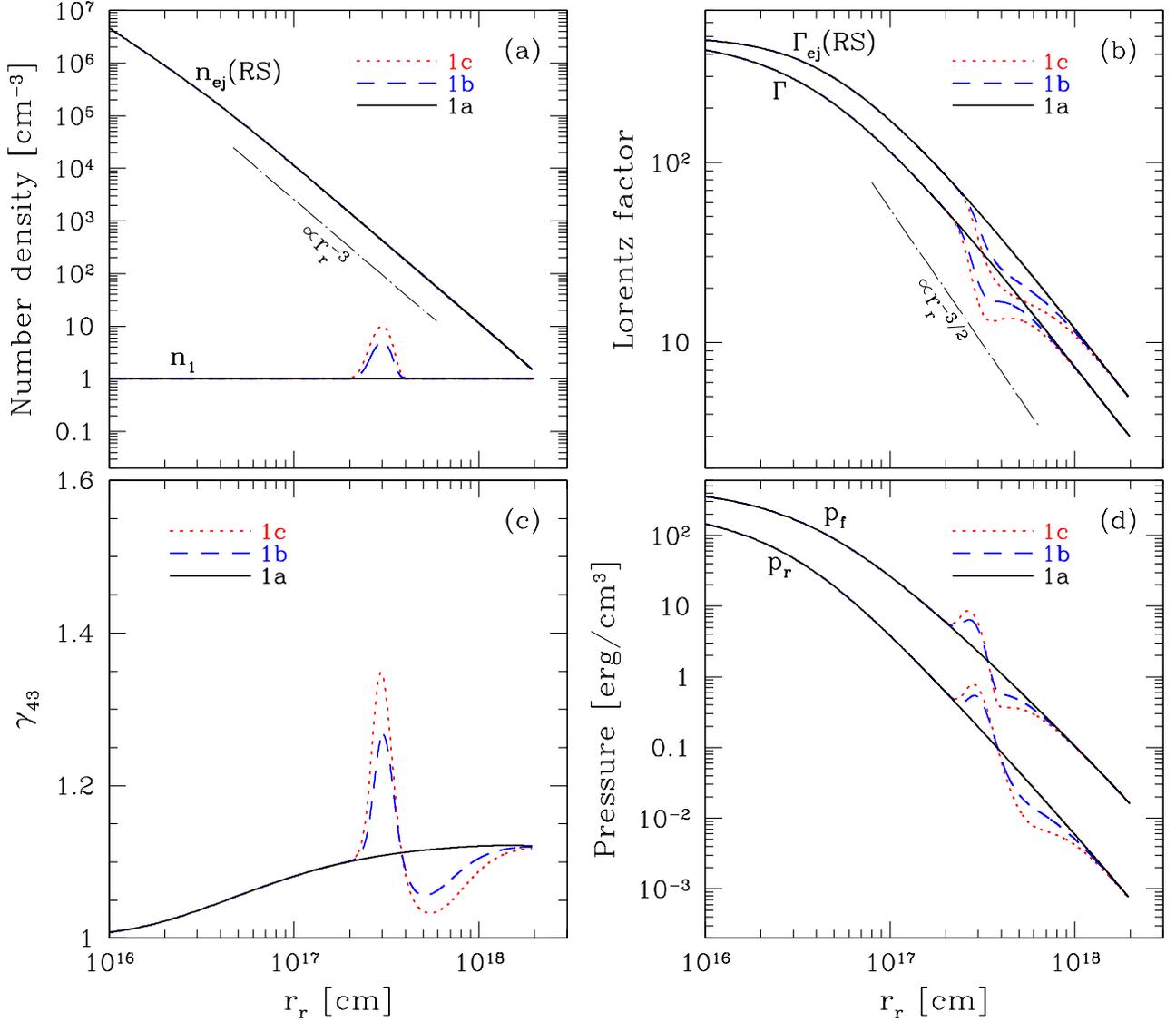}
\caption{
Blast wave dynamics of examples 1a, 1b, and 1c. 
The ambient medium density profile $n_1 = \rho_1/m_p$ of each example is shown in panel (a) 
in different line (color) types. Panel (a) also shows the density $\nejRS = \rhoejRS/m_p$ 
of the ejecta shell that gets shocked by the RS when the RS is located at radius $r_r$. 
In panel (b), three curves labeled by $\GejRS$ show the Lorentz factor of the ejecta shell 
that passes through the RS when the RS is at $r_r$. Three curves labeled by $\Gamma$ show 
the Lorentz factor of the blast wave as a function of the RS radius $r_r$. The relative Lorentz 
factor $\gamma_{43}$ between $\Gamma$ and $\GejRS$ is shown in panel (c). Panel (d) shows 
the evolution of the RS pressure $p_r$ and FS pressure $p_f$ as a function of $r_r$. 
} 
\label{fig:alldyn_1a1b1c}
\end{center}      
\end{figure}

The density profiles of the examples 1a, 1b, and 1c are shown together in the panel (a) 
of Figure~\ref{fig:alldyn_1a1b1c}; three different line (color) types are used to denote 
for each example, respectively. The panel (a) also shows the density $\nejRS = \rhoejRS / m_p$ 
of the ejecta shell that gets shocked by the RS when the RS is located at radius $r_r$. 
For a non-stratified ejecta, 
a simple spherical expansion leads to an $\propto r^{-2}$ decrease in the ejecta density. However, 
for a stratified ejecta, the ejecta density decreases faster than $\propto r^{-2}$ due to a radial 
spread-out of the evolving ejecta flow with a non-vanishing gradient $\Gej^{\prime}(\tau)$. 
In the case of an exponential stratification of $\Gej(\tau)$, U12 (Section 4.1) shows that 
$g_{\rm ej}(\tau) \equiv \left[-\,\frac{d}{d\tau}(\ln \Gej) \right]^{-1}$ has a constant value, and 
as a result, $\nejRS \propto r_r^{-3}$ is satisfied for $g_{\rm ej}(\tau) \ll \frac{r}{c\Gej^2}$. 
This behavior is shown in the panel (a). In the panel (b), the three curves labeled by $\GejRS$ 
show the Lorentz factor of the ejecta shell that passes through the RS when the RS is 
at $r_r$. The three curves labeled by $\Gamma$ show the Lorentz factor of the blast wave 
as a function of the RS radius $r_r$. The relative Lorentz factor $\gamma_{43}$ between $\Gamma$ 
and $\GejRS$ is shown in the panel (c). The $\gamma_{43}$ curve is a measure of the RS strength. 
Panel (d) shows the evolution of the RS pressure $p_r$ and FS pressure $p_f$ as a function of $r_r$.

The ejecta shells with low Lorentz factors run after the blast wave and gradually 
``catch up'' with it at late times, adding their energy to the blast. Thus, 
the blast wave is continuously pushed, and its deceleration deviates from the self-similar 
solution of \cite{blandford76}; the $\Gamma$ curve of the example 1a shows that its blast wave 
decelerates slower than the BM solution $\Gamma \propto r_r^{-3/2}$. For the examples 1b and 1c, 
while encountering a density bump, the $p_f$ curve rises and as a consequence the $\Gamma$ curve 
decreases strongly. Then, the shell catching-up speeds up (as indicated by a strong rise 
in $\gamma_{43}$ curve) and quickly occurs for a wide range of ejecta shells (as shown by a fast 
decrease in $\GejRS$ curve). The rise in $\gamma_{43}$ curve gives the rise in the $p_r$ curve. 
Now climbing down the density bump, the $p_f$ curve decreases rapidly. The $p_f$ curve 
goes even below the solid (black) curve of the example 1a since its blast wave has been 
decelerated significantly due to the bump encountering. The rapid decline in $p_f$ curve, 
together with adiabatic cooling of the blast wave, makes the $\Gamma$ curve start to recover 
from a fast decline phase; the $\GejRS$ curve also starts to flatten, changing its shape 
to a concave curve. Consequently, the shell catching-up slows down and the $\gamma_{43}$ curve 
decreases, so does the $p_r$ curve.

Moving away from the bump region, the $p_f$ curve is still well below the solid (black) curve 
of the example 1a, retaining in memory the previous deceleration history due to the bump. 
The low FS pressure, combined with ongoing adiabatic cooling of the blast wave, makes 
the $\Gamma$ curve continue to recover; note that the $\Gamma$ curve of the example 1c 
exhibits a period of an actual acceleration of the blast wave. This continuing recovery 
in $\Gamma$ curve further slows down the shell catching-up and makes the $\gamma_{43}$ curve 
continue to decrease below the solid (black) curve. Accordingly, the $p_r$ curve also goes 
below the solid (black) curve. As the $p_f$ curve approaches back to the solid (black) curve, 
the recovery tendency in $\Gamma$ curve weakens and the $\gamma_{43}$ curve begins to rise again. 
Since the examples 1a, 1b, and 1c are given the same burst energy and ejecta stratification, 
the three $\Gamma$ curves, after all, should agree with one another,\footnote{Although the blast 
waves of these three examples carry a different amount of mass in the bumps, this mass 
difference becomes negligible as the blast waves propagate further away from the bump location, 
since the mass contained in the bumps will become negligibly small when compared to the total 
swept-up mass.} 
if the adiabatic conversion of internal energy has been properly taken care of. 
An important consistency test of whether the modeling is correct would be 
whether different models converge after the bump phase. An excellent agreement 
in $\Gamma$ curves as well as in other curves (i.e., $\GejRS$, $\gamma_{43}$, $p_f$, 
and $p_r$ curves) is seen in Figure~\ref{fig:alldyn_1a1b1c}, suggesting the correctness 
of the modeling. 
Also, as described by the example 1c and other examples below, 
correct modeling should be able to capture an acceleration phase of the blast wave 
when the ambient medium density drops rapidly.
\cite{nava13} also considered the role of adiabatic cooling 
and reproduced an acceleration phase of the blast wave.

\begin{figure}
\begin{center}
\includegraphics[width=17cm]{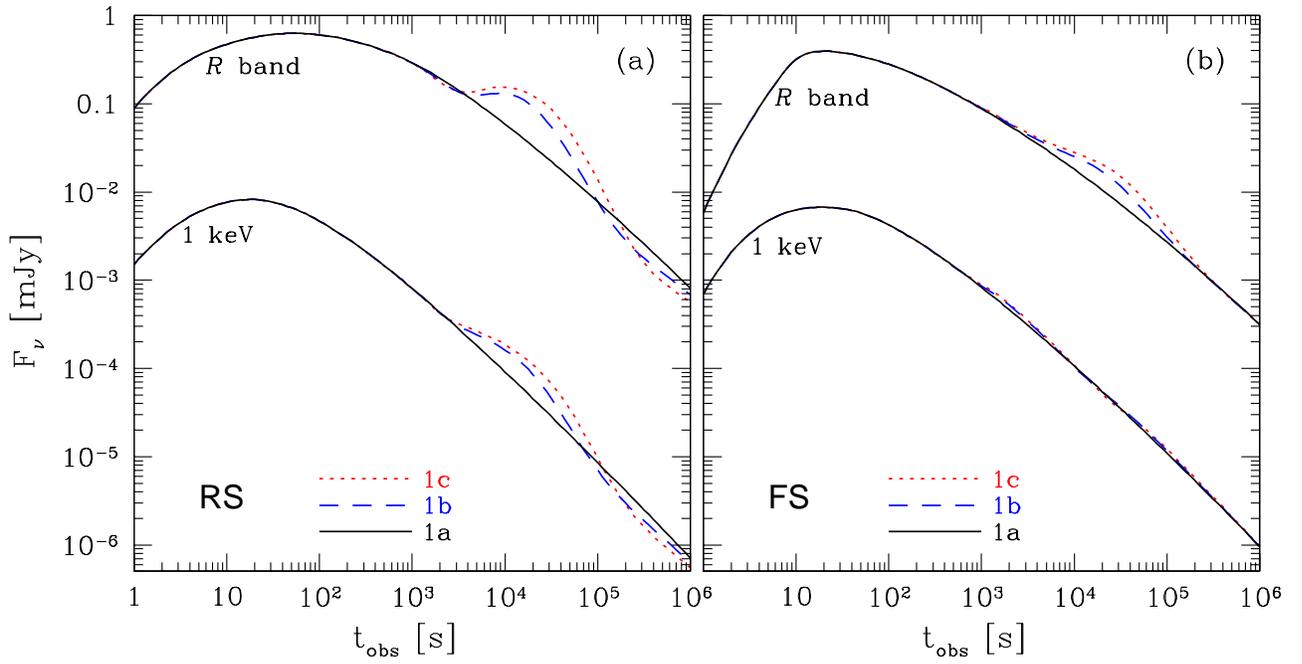}
\caption{
Afterglow light curves of examples 1a, 1b, and 1c. 
Panel (a) shows the RS emission in X-ray (1 keV) and {\it R} band as a function of the 
observer time $\tobs$. Panel (b) shows the FS emission in X-ray (1 keV) and {\it R} band. 
} 
\label{fig:cRX_1a1b1c}
\end{center}      
\end{figure}

The afterglow light curves of the examples 1a, 1b, and 1c are shown in Figure~\ref{fig:cRX_1a1b1c}. 
In panel (a), we show the RS emission in X-ray (1 keV) and {\it R} band as a function of the 
observer time $\tobs$. In panel (b), we show the FS emission in X-ray (1 keV) and {\it R} band. 
The example 1a shows that, during the deceleration phase, the FS and RS light curves decline with 
a very similar value of decay indices; a detailed explanation regarding this point is given 
in U12. For the examples 1b and 1c, while encountering a density bump, a signature corresponding to 
the bump is visible on both the FS and the RS light curves, but with a stronger re-brightening 
feature imprinted on the RS light curves. The feature is stronger for the example 1c than for 
the example 1b, since the bump is bigger in the example 1c than in the example 1b. 
After exhibiting the re-brightening feature, the RS light curves go even below the solid (black) 
curve of the example 1a and then approach back to the solid (black) curve (just as the $\gamma_{43}$ 
curve behaves). This interesting behavior is not clearly present in the FS light curves.

\subsection{Group (1a/1c/1d)}

This group includes a new example 1d, whose density profile $n_1(r)$ in the ambient medium also 
has a Gaussian bump on top of the base density $n_0=1~\mbox{cm}^{-3}$ at the same location 
$r_0=3 \times 10^{17}~\mbox{cm}$. The amplitude of the bump is $n_a=19~\mbox{cm}^{-3}$ so that 
the peak density is 20 times the base density. As for the spatial spread of the bump, 
we choose $\sigma$ such that the bump of the examples 1c and 1d contains a same amount of mass, 
$\int_0^\infty m_p \left[n_1(r)-n_0 \right] 4\pi r^2\,{\rm d}r$. Recalling that the example 1c 
has $n_a=9~\mbox{cm}^{-3}$ and $\sigma=\sigma_0=3 \times 10^{16}~\mbox{cm}$, we obtain $\sigma$ 
for the example 1d, which is about $0.477\,\sigma_0$.

\begin{figure}
\begin{center}
\includegraphics[width=17cm]{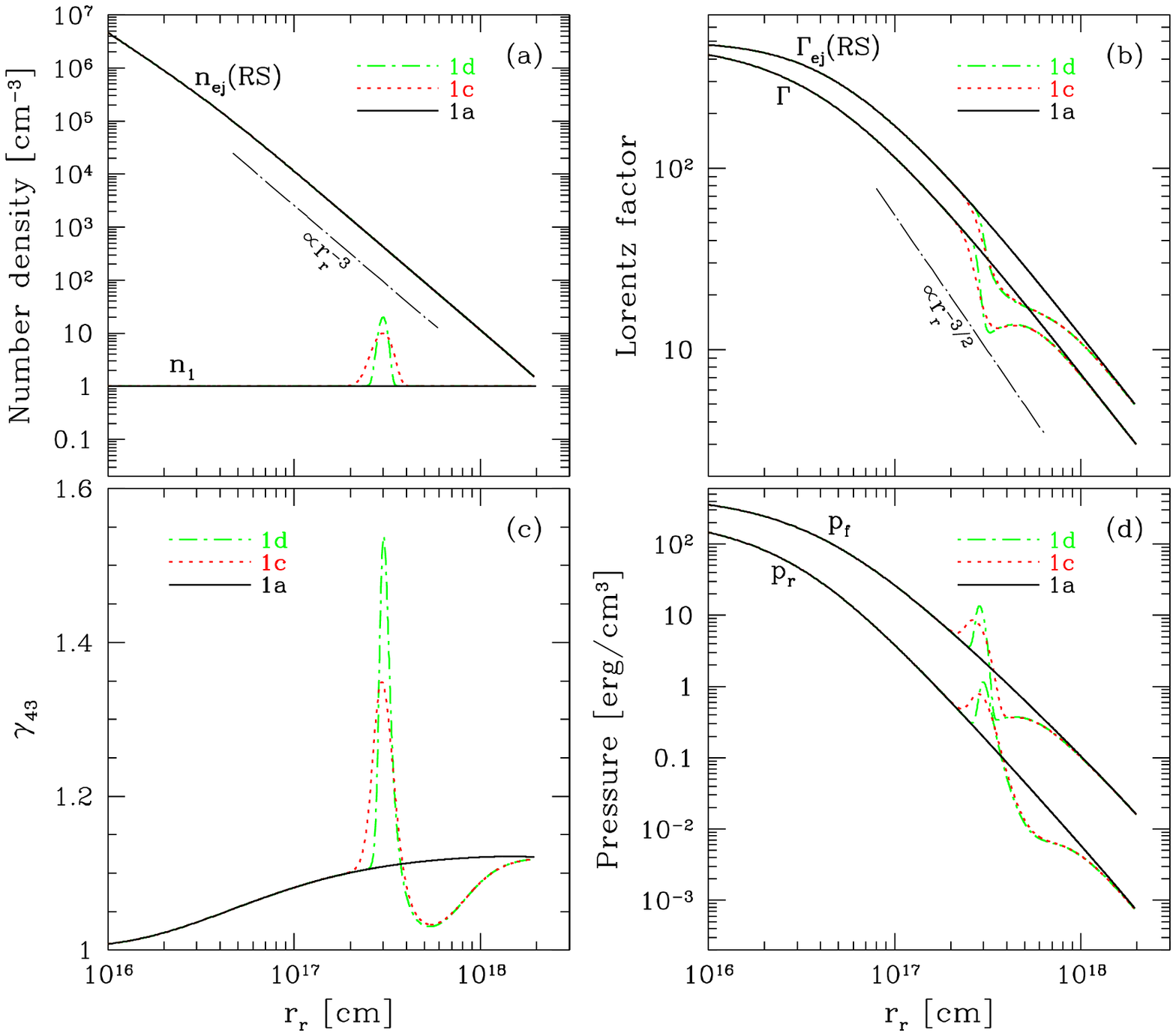}
\caption{
Same as in Figure~\ref{fig:alldyn_1a1b1c}, but for examples 1a, 1c, and 1d. 
} 
\label{fig:alldyn_1a1c1d}
\end{center}      
\end{figure}

The blast wave dynamics of the examples 1a, 1c, and 1d are shown together 
in Figure~\ref{fig:alldyn_1a1c1d}; all 4 panels have the same notations as in the previous 
group (1a/1b/1c). One can read through the panels and understand the FS and RS dynamics, in 
the same way as we did above. In particular, we point out that since the examples 1c and 1d 
have the same mass contained in the bump, their dynamical evolution (i.e., $\Gamma$, $\GejRS$, 
$\gamma_{43}$, $p_f$, and $p_r$ curves) agrees with each other as soon as their blast wave goes 
beyond the bump region (even before converging with the solid (black) curves of the example 1a). 
This agreement among examples with an equal bump mass serves as another consistency test.

\begin{figure}
\begin{center}
\includegraphics[width=17cm]{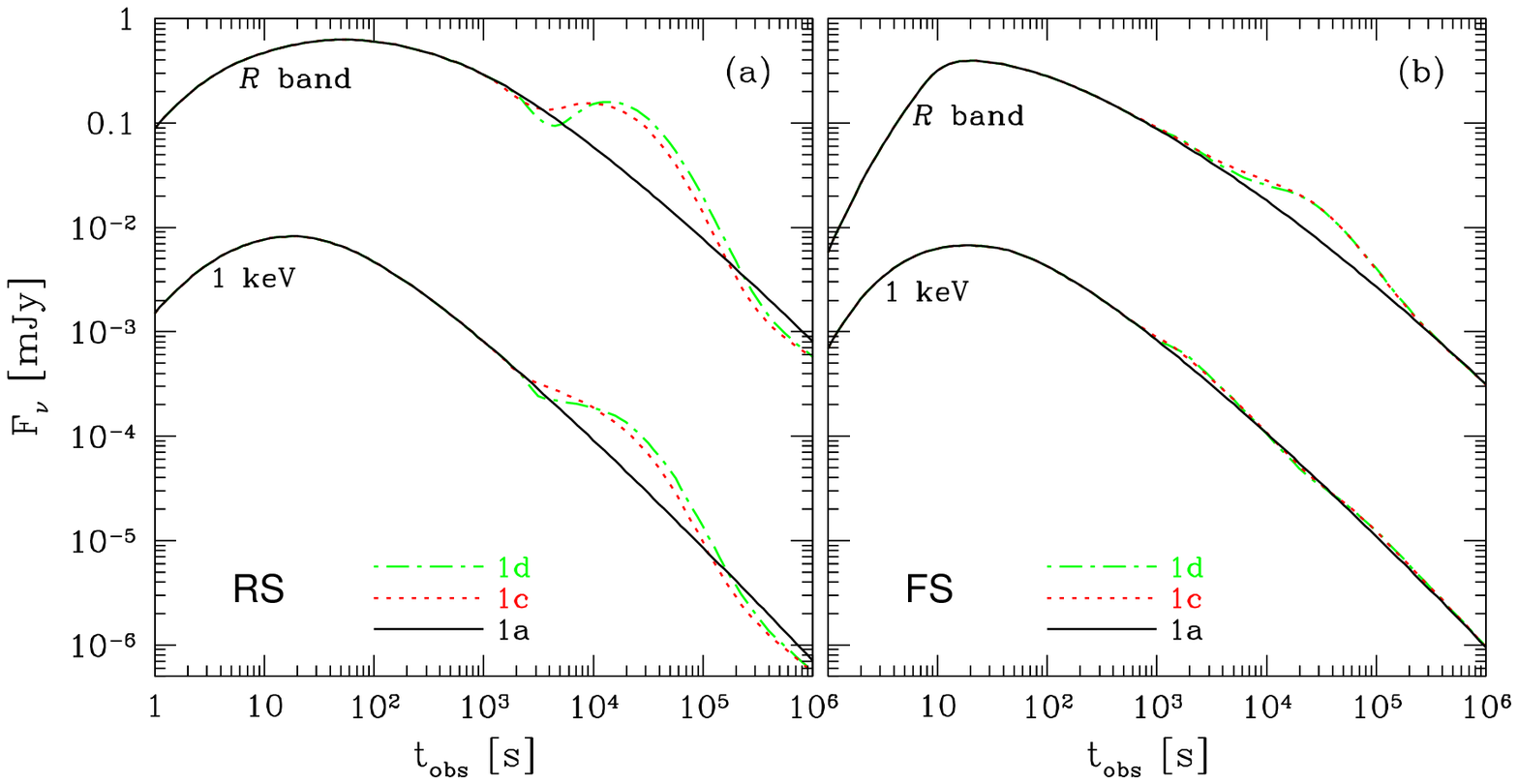}
\caption{
Same as in Figure~\ref{fig:cRX_1a1b1c}, but for examples 1a, 1c, and 1d. 
} 
\label{fig:cRX_1a1c1d}
\end{center}      
\end{figure}

The afterglow light curves of the examples 1a, 1c, and 1d are shown together 
in Figure~\ref{fig:cRX_1a1c1d}. 
Although the examples 1c and 1d have the same bump mass, the bump in the example 1d is narrower 
than in the example 1c. Accordingly, the RS light curves show a more concentrated re-brightening 
feature in the example 1d than in the example 1c. For the FS light curves, the difference between 
the two examples appears to be very small. 

\subsection{Group (1a/1e/1f)}

The new examples 1e and 1f here have a density void in the ambient medium, located at radius $r_0$. 
A Gaussian shape is removed from our base density $n_0$ and thus the density profile is given 
by $n_1(r)=n_0-n_a\, e^{-(r-r_0)^2/(2\sigma^2)}$. For both examples 1e and 1f, we use a void amplitude 
$n_a=0.9~\mbox{cm}^{-3}$ so that the minimum density is 1/10 of the base density. As for the spatial 
spread of the void, the example 1e has $\sigma=\sigma_0$ while the example 1f has $\sigma=2\,\sigma_0$.

\begin{figure}
\begin{center}
\includegraphics[width=17cm]{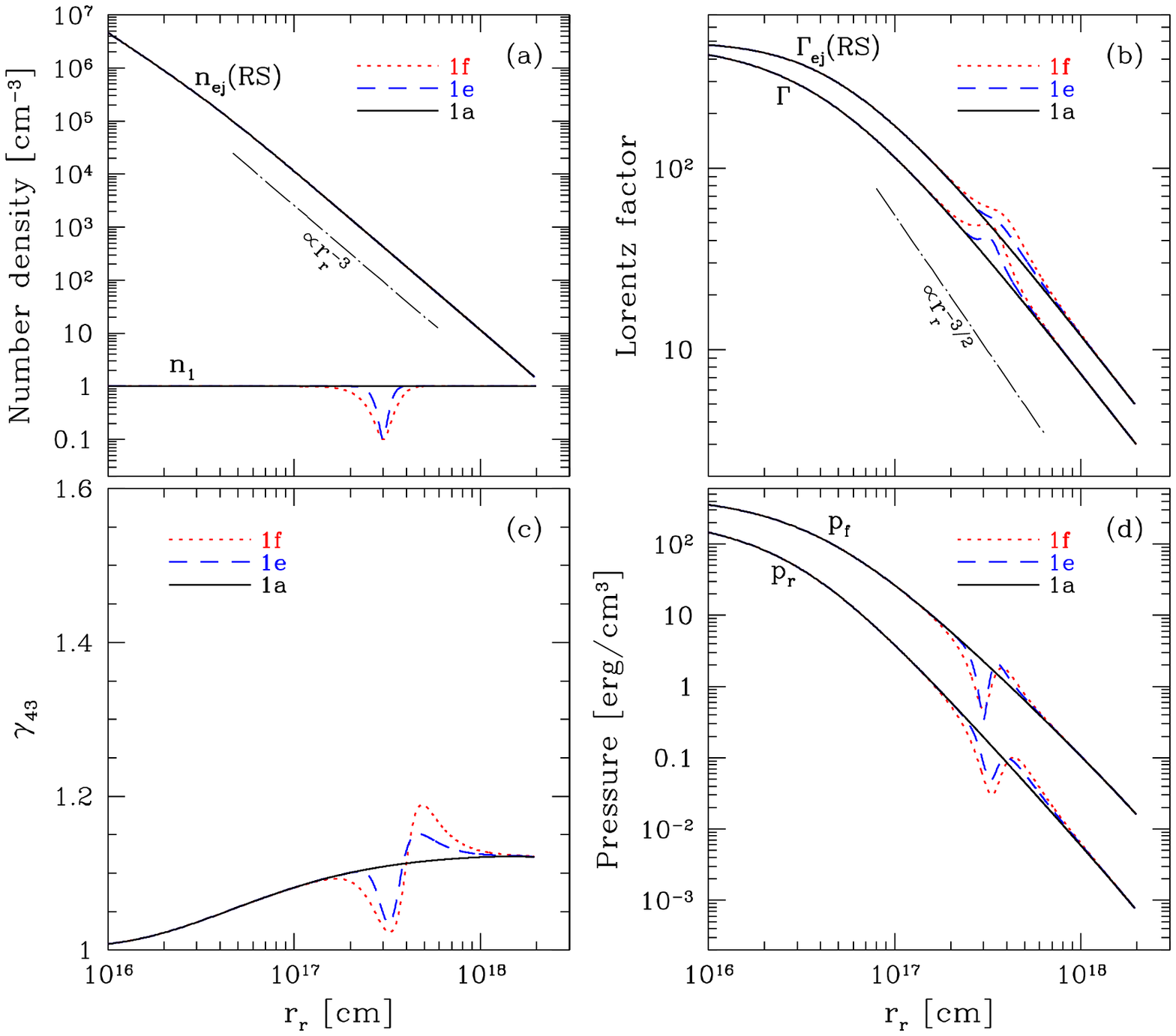}
\caption{
Same as in Figure~\ref{fig:alldyn_1a1b1c}, but for examples 1a, 1e, and 1f. 
} 
\label{fig:alldyn_1a1e1f}
\end{center}      
\end{figure}

The blast wave dynamics of the examples 1a, 1e, and 1f are shown together 
in Figure~\ref{fig:alldyn_1a1e1f}. When the blast wave of the examples 1e and 1f encounters a density 
void shown in panel (a), its $p_f$ curve declines rapidly. 
This $p_f$ decline then, together with adiabatic cooling of the blast wave, 
makes the blast wave decelerate slowly, resulting in a flattening in the $\Gamma$ curve. 
As a result, the shell catching-up occurs slowly (as indicated by a flattening in the $\GejRS$ curve 
and also by a decline in the $\gamma_{43}$ curve). The decline in the $\gamma_{43}$ curve gives the decline 
in the $p_r$ curve. Now climbing up the void hill (the outer side of void), the $p_f$ curve rises rapidly. 
The $p_f$ curve goes even above the solid (black) curve of the example 1a since its $\Gamma$ curve is 
above the solid (black) curve (due to the void encountering). The rapid rise in $p_f$ curve makes 
the $\Gamma$ curve start to turn into a fast decline phase; the $\GejRS$ curve also changes its shape 
to a convex curve. Consequently, the shell catching-up speeds up and the $\gamma_{43}$ curve resumes
rising. The $p_r$ curve also rises back as the $\gamma_{43}$ curve resumes rising.

Escaping from the void region, the $p_f$ curve is still above the solid (black) curve of the 
example 1a, retaining the memory of the previous history of void encountering. The high FS pressure 
makes the $\Gamma$ curve continue to decline fast and thus makes the $\gamma_{43}$ curve continue 
to rise above the solid (black) curve. Hence, the $p_r$ curve also goes above the solid (black) curve. 
As the $p_f$ curve gradually approaches back to the solid (black) curve, the fast-decline tendency 
in $\Gamma$ curve weakens and therefore the $\gamma_{43}$ curve begins to decrease again. 
For the same reason we discussed above for bumps, the blast wave evolution (i.e., $\Gamma$, 
$\GejRS$, $\gamma_{43}$, $p_f$, and $p_r$ curves) of these three examples 1a, 1e, and 1f should 
agree with one another, after all. The agreement shown in Figure~\ref{fig:alldyn_1a1e1f}, 
while invoking for density voids, also serves as a consistency test for the correctness of modeling.

\begin{figure}
\begin{center}
\includegraphics[width=17cm]{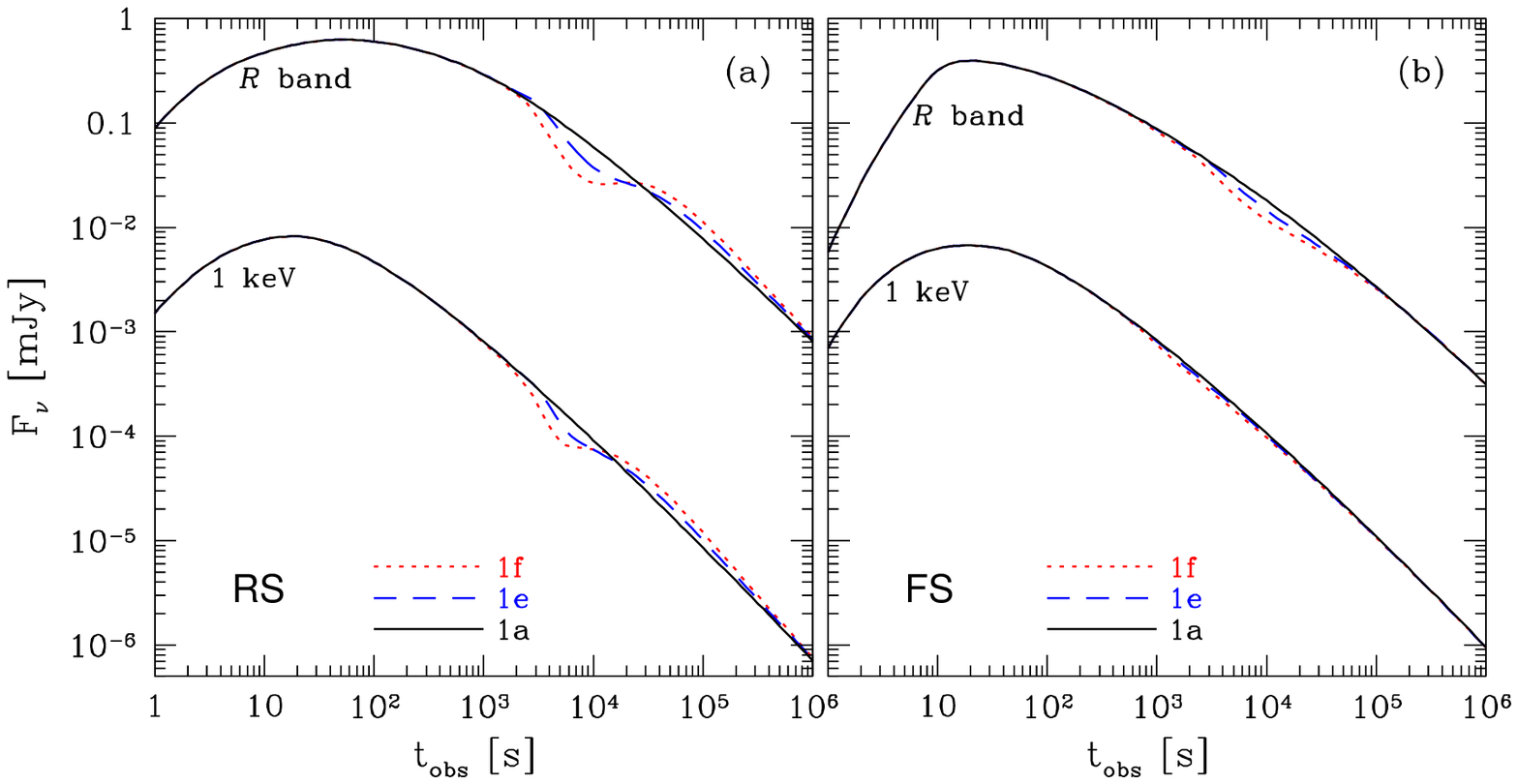}
\caption{
Same as in Figure~\ref{fig:cRX_1a1b1c}, but for examples 1a, 1e, and 1f. 
} 
\label{fig:cRX_1a1e1f}
\end{center}      
\end{figure}

The afterglow light curves of the examples 1a, 1e, and 1f are shown in Figure~\ref{fig:cRX_1a1e1f}. 
For the examples 1e and 1f, while encountering a density void, a dip signature corresponding to 
the void is visible on both the FS and the RS light curves, but with a stronger feature 
imprinted on the RS light curves. The feature is stronger for the example 1f than for 
the example 1e, since the void is bigger in the example 1f than in the example 1e. 
After exhibiting the dip feature, the RS light curves go even above the solid (black) 
curve of the example 1a and then approach back to the solid (black) curve, just as 
the $\gamma_{43}$ curve behaves. This behavior is not present in the FS light curves.

\subsection{Group (1a/1f/1g)}

The new example 1g here has a density void in the ambient medium at the same location $r_0$. 
Its density profile is given by $n_1(r)=n_0-n_a\, e^{-(r-r_0)^4/(4\sigma^4)}$, 
with the void amplitude $n_a=0.95~\mbox{cm}^{-3}$ so that the minimum density is 1/20 of 
the base density $n_0$. We choose $\sigma$ such that the examples 1f and 1g 
have a same value for the integration 
$\int_0^\infty m_p \left[n_1(r)-n_0 \right] 4\pi r^2\,{\rm d}r$, to ensure that an equal amount 
of mass is reduced by the void in the examples 1f and 1g. Note that we use a boxier shape of 
functional form for the void of the example 1g, so that an equality of the integration above 
gives a narrower void for the example 1g.

\begin{figure}
\begin{center}
\includegraphics[width=17cm]{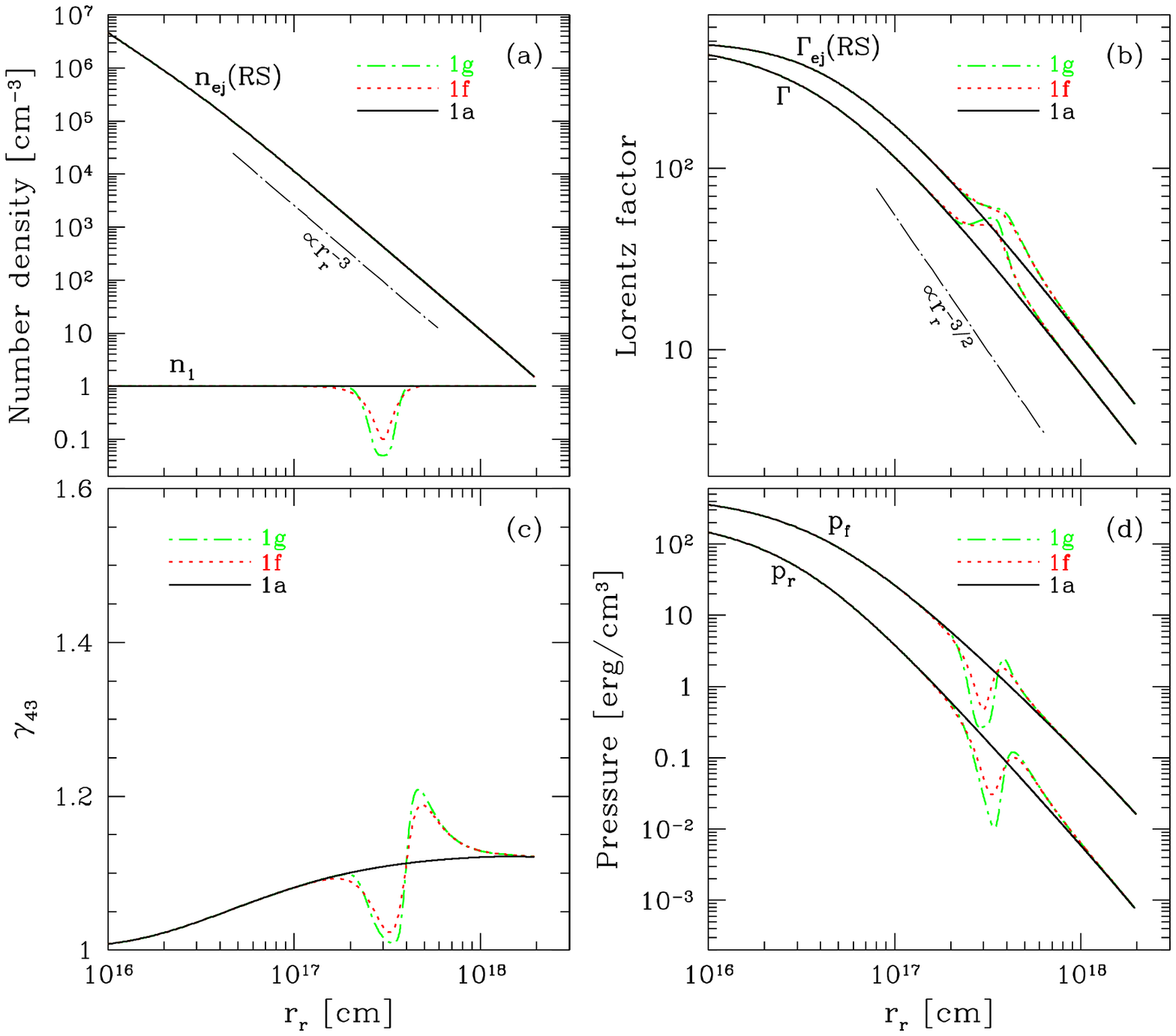}
\caption{
Same as in Figure~\ref{fig:alldyn_1a1b1c}, but for examples 1a, 1f, and 1g. 
} 
\label{fig:alldyn_1a1f1g}
\end{center}      
\end{figure}

The blast wave dynamics of the examples 1a, 1f, and 1g are shown together 
in Figure~\ref{fig:alldyn_1a1f1g}. Again, one can read through the panels and understand 
the FS and RS dynamics. Note that the $\Gamma$ curve of the example 1g shows a period 
of an actual acceleration of the blast wave. Since the void of the examples 1f and 1g 
reduces an equal amount of mass from the base, the blast wave evolution of these two 
examples agrees with each other as soon as their blast wave goes beyond the void region, 
even before converging with the solid (black) curves of the example 1a. 
This serves as another consistency test.

\begin{figure}
\begin{center}
\includegraphics[width=17cm]{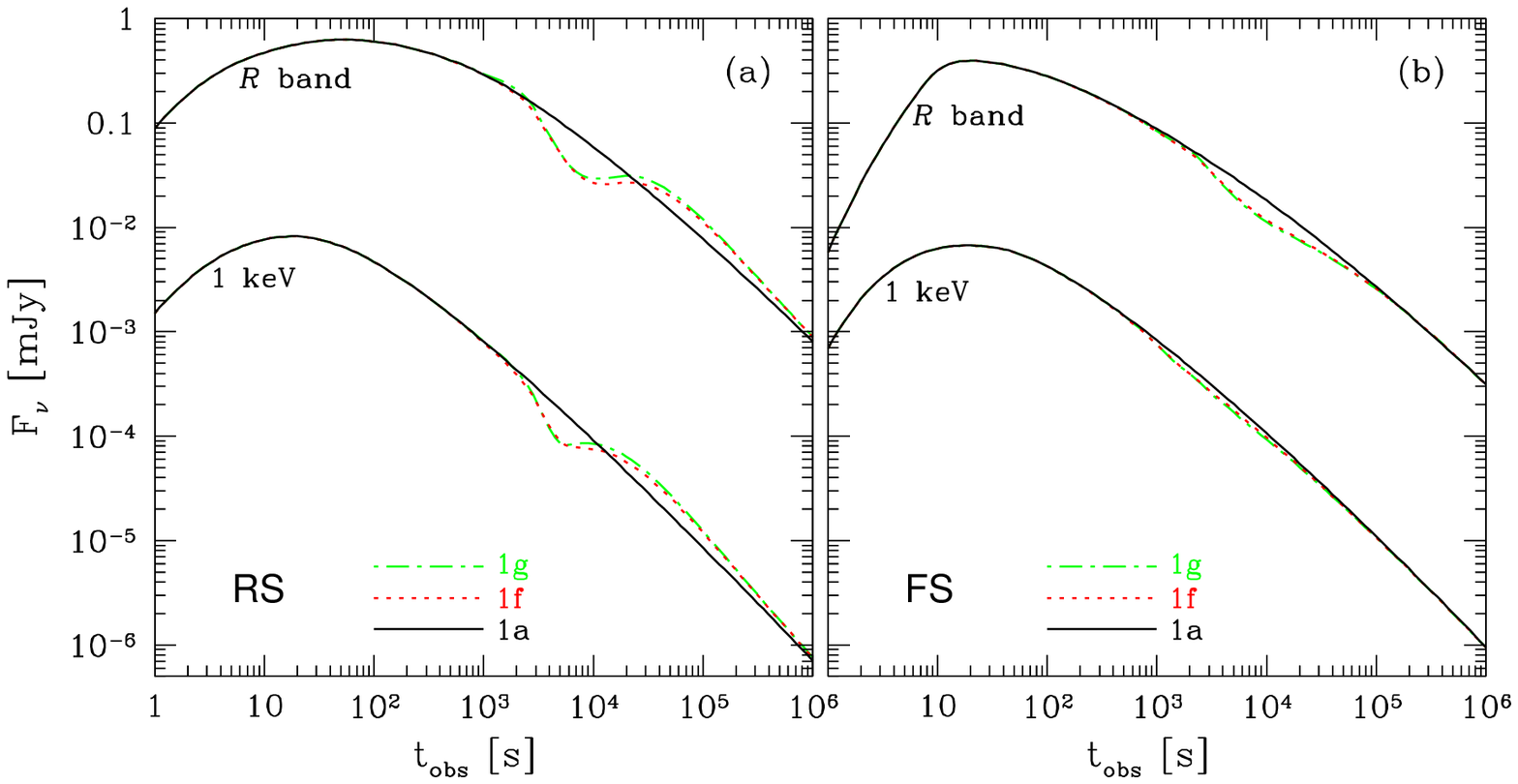}
\caption{
Same as in Figure~\ref{fig:cRX_1a1b1c}, but for examples 1a, 1f, and 1g. 
} 
\label{fig:cRX_1a1f1g}
\end{center}      
\end{figure}

The afterglow light curves of the examples 1a, 1f, and 1g are shown together 
in Figure~\ref{fig:cRX_1a1f1g}. The examples 1f and 1g show a small difference 
on the RS light curves, while almost no difference on the FS light curves is 
observed between these two examples.

\subsection{Group (1a/1h/1i)}

\begin{figure}
\begin{center}
\includegraphics[width=17cm]{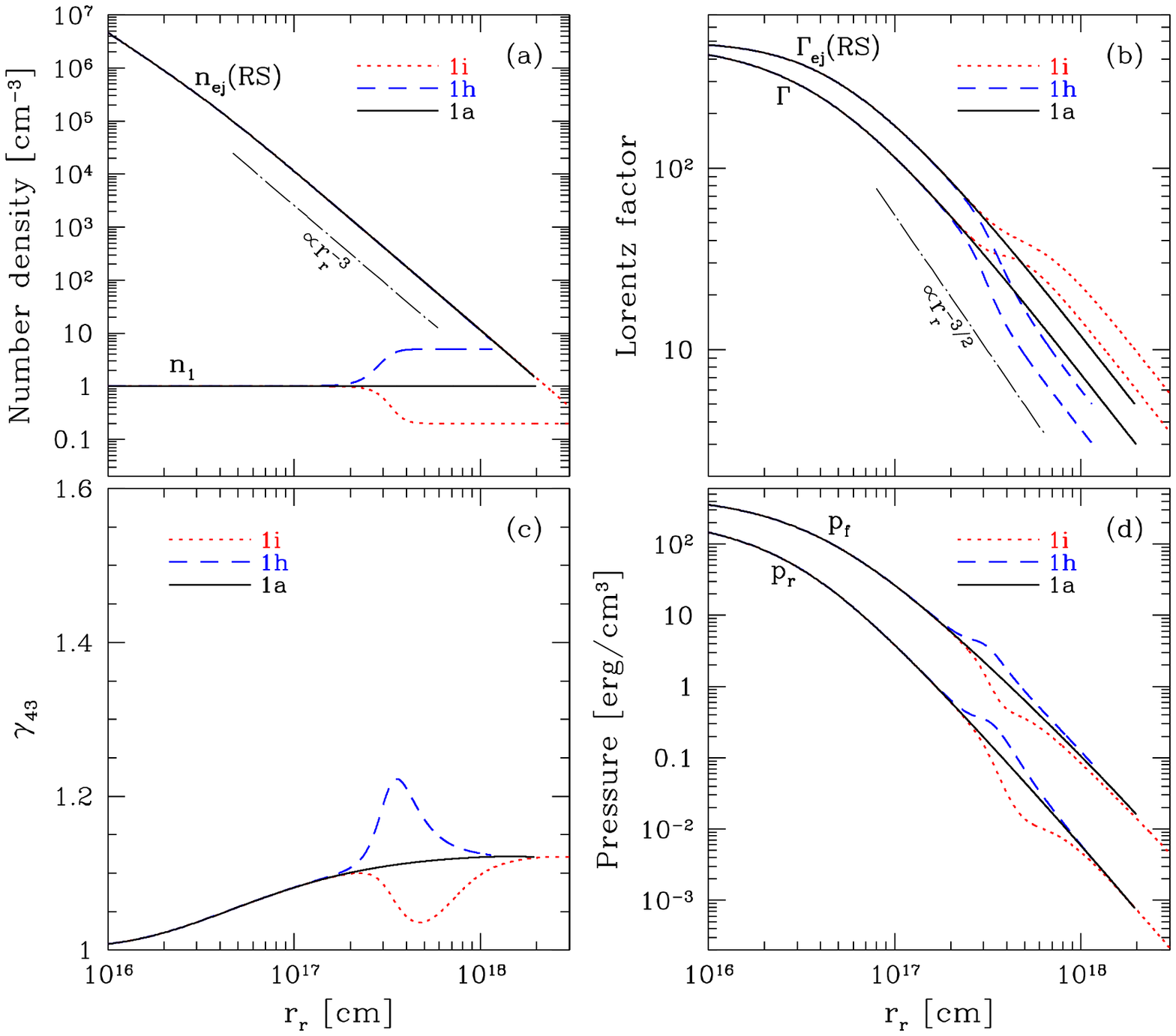}
\caption{
Same as in Figure~\ref{fig:alldyn_1a1b1c}, but for examples 1a, 1h, and 1i. 
} 
\label{fig:alldyn_1a1h1i}
\end{center}      
\end{figure}

The new examples 1h and 1i invoke a step-like density structure. 
The example 1h has a density profile 
$n_1(r)=n_0+n_a\,[1+e^{-(r-r_0)/\sigma_0}]^{-1}$, 
with an amplitude $n_a=4~\mbox{cm}^{-3}$. The density increases 
from the base $n_0$ by a factor of 5, through the step that is located 
at a radius $r_0$ in a length scale $\sigma_0$. The example 1i has the same 
density profile except that $n_a=-0.8~\mbox{cm}^{-3}$. Then, the density 
decreases from the base by a factor of 5. 
See panel (a) in Figure~\ref{fig:alldyn_1a1h1i}.

The blast wave dynamics of the examples 1a, 1h, and 1i are shown together 
in Figure~\ref{fig:alldyn_1a1h1i}. For the example 1h, while encountering the rising step, 
the blast wave shows the same behavior as in the rising phase of the bumps. However, 
after that there is no rapid decline in the density and thus no rapid decline in the $p_f$ 
curve. Therefore, the $\Gamma$ curve does not recover from the fast decline and stay well 
below the solid (black) curve. Propagating further on a constant density medium lifted up 
from the base, the blast wave gradually stabilizes and the $\gamma_{43}$ curve returns 
back to the solid (black) curve, so does the $p_r$ curve. 
The $p_f$ curve also approaches back to the solid (black) curve (propagating with lower $\Gamma$ 
at higher $n_1$ region), but does not agree with the solid (black) curve. 
Similarly, for the example 1i, while encountering the descending step, the blast wave shows 
the same behavior as in the descending phase of the voids. Again, after that there is no rapid 
rise in density and thus no rapid rise in the $p_f$ curve. Hence, the $\Gamma$ curve does 
not turn into a rapid decline phase and stay well above the solid (black) curve. Moving further 
on a constant density medium shifted down from the base, the blast wave gradually adjusts to 
the lower step and the $\gamma_{43}$ curve returns to the solid (black) curve, so does the $p_r$ 
curve. The $p_f$ curve also approaches back to the solid (black) curve (propagating with 
higher $\Gamma$ at lower $n_1$ region), but does not completely agree with the solid (black) curve.

\begin{figure}
\begin{center}
\includegraphics[width=17cm]{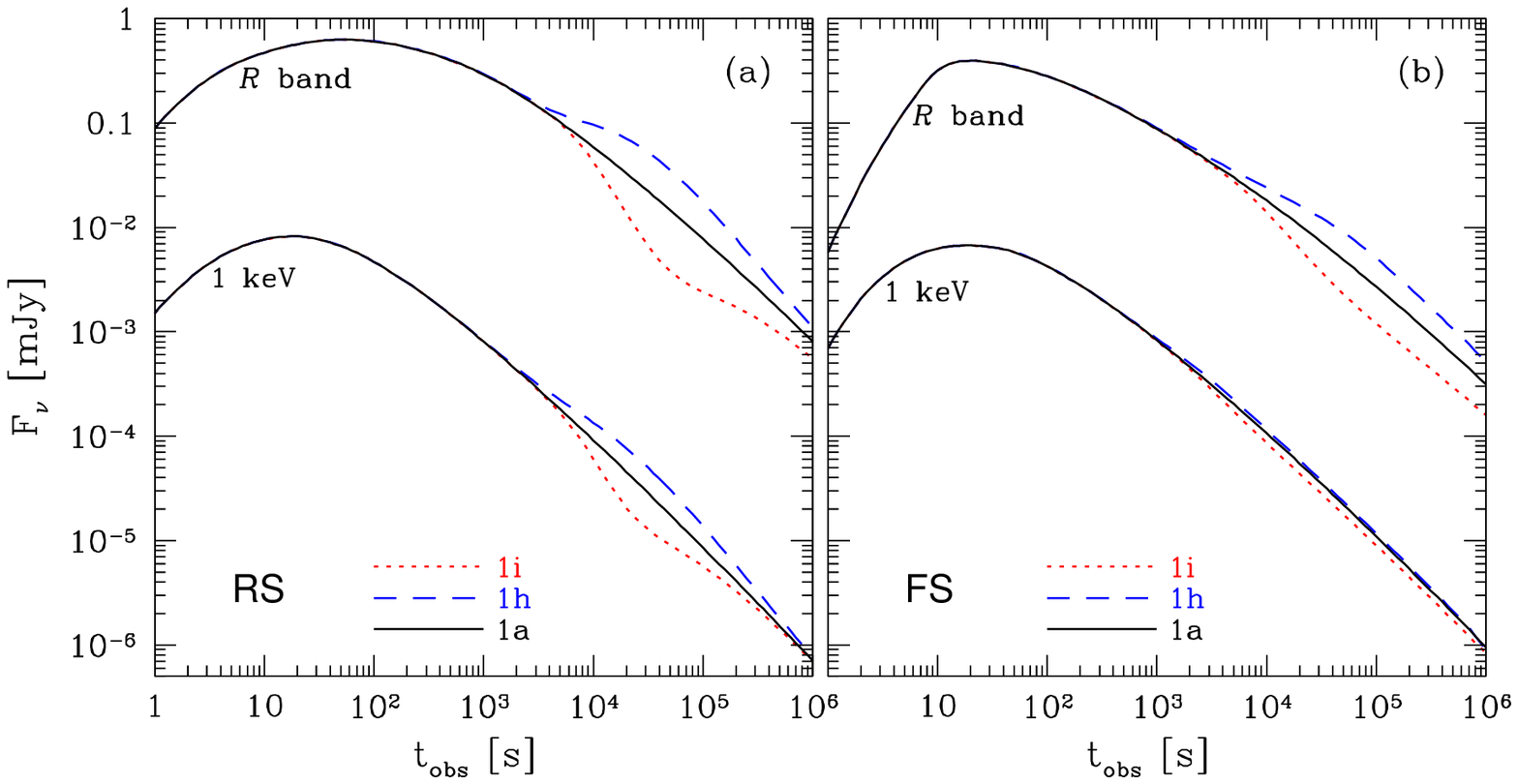}
\caption{
Same as in Figure~\ref{fig:cRX_1a1b1c}, but for examples 1a, 1h, and 1i. 
} 
\label{fig:cRX_1a1h1i}
\end{center}      
\end{figure}

The afterglow light curves of the examples 1a, 1h, and 1i are shown together 
in Figure~\ref{fig:cRX_1a1h1i}. For the example 1h, the RS light curves show a mild re-brightening 
and then approach back to the solid (black) curve. The FS light 
curves also respond to the rising phase of the step structure, but do not approach back to the solid 
(black) curve. For the example 1i, the RS light curves decline initially, responding to the descending 
phase of the step structure, and then go back to the solid (black) curve. 
Note that this kind of behavior in light curves has often been interpreted 
as an FS wave exhibiting a jet break (for the initial steepening), followed by a late-time energy 
injection (for the flattening). Here we present a simple way to reproduce such a behavior, i.e., 
an RS wave responding to a descending step structure. The FS light curves of the example 1i also 
respond to the step, but again do not approach back to the solid (black) curve.

\subsection{Group (2a/2c/2d)}

As mentioned above, the examples with the number 2 in their names have a short-lived RS, 
with a constant ejecta Lorentz factor $\Gej=300$. For the external ambient medium, three 
examples 2a, 2c, and 2d here share the same density profile with the examples 1a, 1c, and 1d, 
respectively. Recall that the example 1a has a constant density $n_0$ (our base), 
and the examples 1c and 1d have a density bump on top of the base.

\begin{figure}
\begin{center}
\includegraphics[width=10cm]{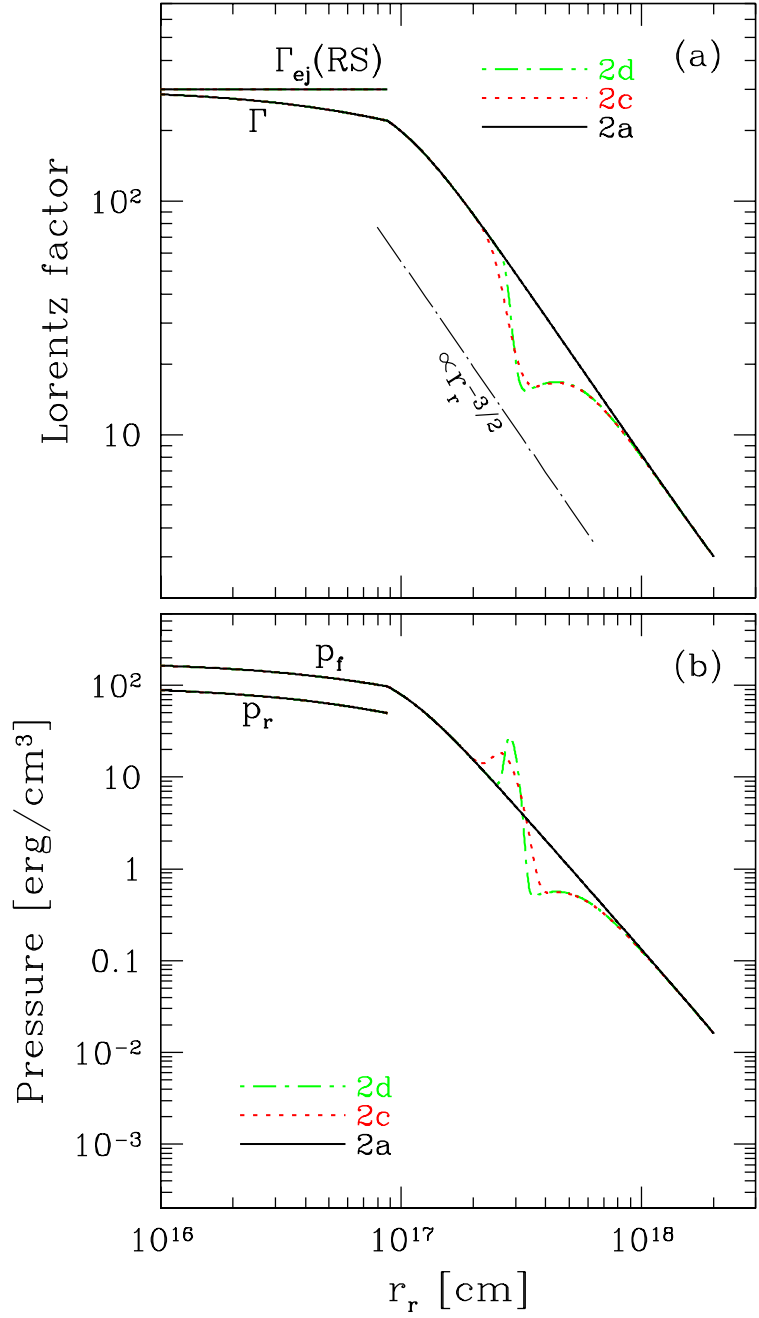}
\caption{
Blast wave dynamics of examples 2a, 2c, and 2d. 
These examples have a short-lived RS. 
The same notations as in Figure~\ref{fig:alldyn_1a1b1c} are used here. 
Panel (a) shows the $\Gamma$ and $\GejRS$ curves, and the panel (b) shows 
the $p_f$ and $p_r$ curves. 
} 
\label{fig:dyn_2a2c2d}
\end{center}      
\end{figure}

In Figure~\ref{fig:dyn_2a2c2d}, we show the blast wave dynamics of the examples 2a, 2c, and 2d 
together. The notations are the same as in the previous groups of examples with a long-lived RS. 
We show the $\Gamma$ and $\GejRS$ curves in the panel (a) and the $p_f$ and $p_r$ curves in the panel (b). 
The $\GejRS$ and $p_r$ curves exhibit an early end when the RS crosses the end of the ejecta. 
After that, there is no more ejecta shells catching up with the blast wave. Hence, the $\Gamma$ curve 
of the example 2a transitions to and then agrees with the Blandford-McKee deceleration. 
For the examples 2c and 2d, the FS dynamics (i.e., $\Gamma$ and $p_f$ curves) responding to 
the density bump can be understood in the same way as in the group (1a/1c/1d). An early agreement 
between 2c and 2d curves beyond the bump region is observed again since the bump in those two 
examples contains a same amount of mass. It is also observed that all three examples 2a, 2c, and 2d 
agree with one another in the FS curves as their blast wave propagates further away from the bump region,
suggesting the correctness of our modeling. These agreements serve as a consistency test here.

\begin{figure}
\begin{center}
\includegraphics[width=10cm]{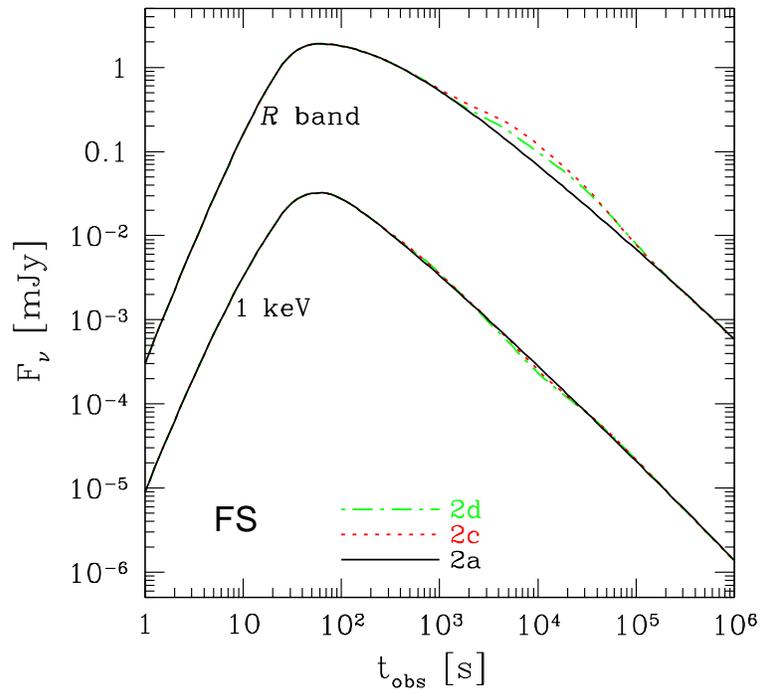}
\caption{
Afterglow light curves of examples 2a, 2c, and 2d. 
The FS emission in X-ray (1 keV) and {\it R} band is shown 
as a function of the observer time $\tobs$. 
} 
\label{fig:cRX_fs_2a2c2d}
\end{center}      
\end{figure}

In Figure~\ref{fig:cRX_fs_2a2c2d}, we show the FS light curves of the examples 2a, 2c, and 2d 
together. For the examples 2c and 2d, the FS light curves responding to the density bump exhibit 
the same behavior as in the group (1a/1c/1d).

\subsection{Group (2a/2f/2g)}

\begin{figure}
\begin{center}
\includegraphics[width=10cm]{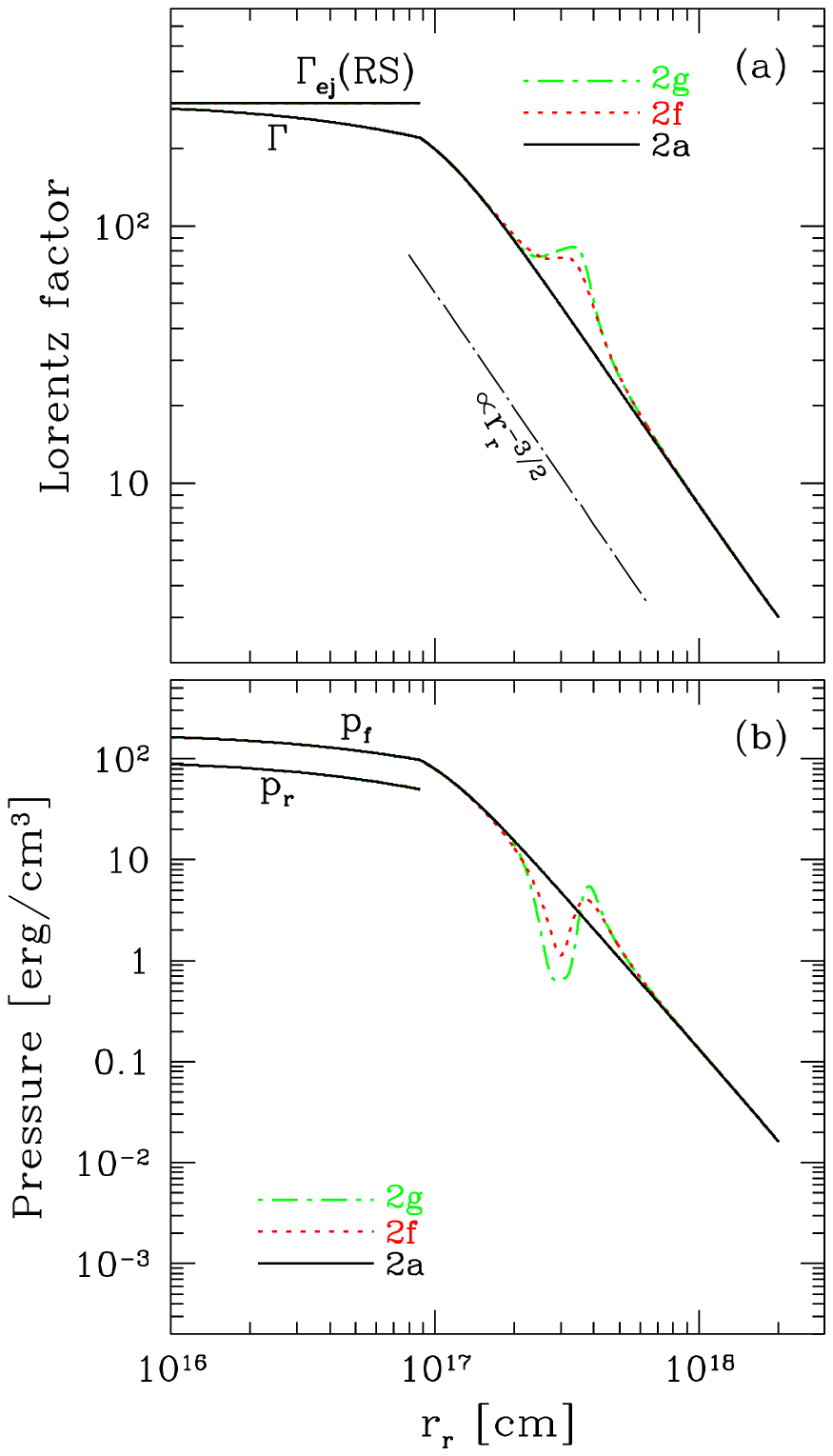}
\caption{
Same as in Figure~\ref{fig:dyn_2a2c2d}, but for examples 2a, 2f, and 2g. 
} 
\label{fig:dyn_2a2f2g}
\end{center}      
\end{figure}

The new examples 2f and 2g share the same density profile with the examples 1f and 1g, 
respectively. Recall that the examples 1f and 1g have a density void in the ambient medium. 
In Figure~\ref{fig:dyn_2a2f2g}, we show the blast wave dynamics of the examples 2a, 2f, and 2g 
together. For the examples 2f and 2g, the FS dynamics (i.e., $\Gamma$ and $p_f$ curves) 
responding to the density void can be understood in the same way as in the group (1a/1f/1g). 
An early agreement between 2f and 2g curves beyond the void region is observed since the void 
in those two examples reduces a same amount of mass from the base. Also, all three examples 
2a, 2f, and 2g agree with one another in the FS curves as their blast wave propagates 
further away from the void region, suggesting the correctness of the modeling.

\begin{figure}
\begin{center}
\includegraphics[width=10cm]{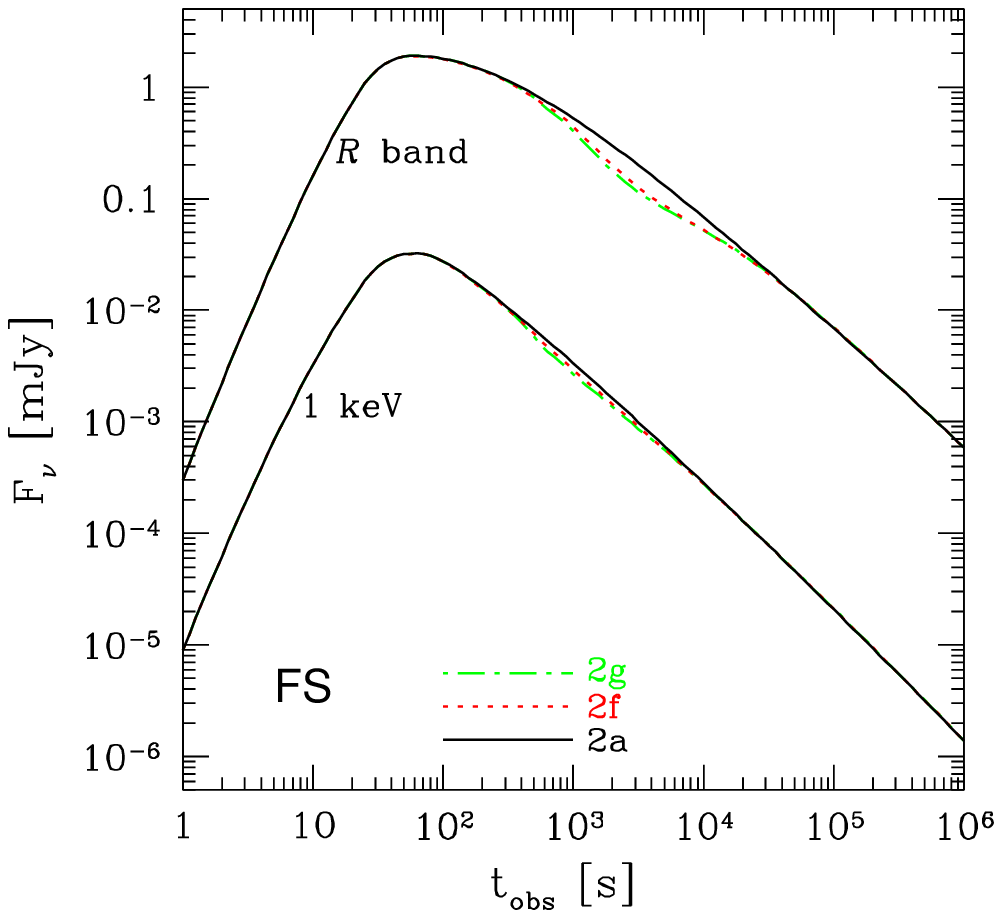}
\caption{
Same as in Figure~\ref{fig:cRX_fs_2a2c2d}, but for examples 2a, 2f, and 2g. 
} 
\label{fig:cRX_fs_2a2f2g}
\end{center}      
\end{figure}

In Figure~\ref{fig:cRX_fs_2a2f2g}, we show the FS light curves of the examples 2a, 2f, and 2g 
together. For the examples 2f and 2g, the FS light curves responding to the density void exhibit 
the same behavior as in the group (1a/1f/1g).

%
%

\section{Conclusions and Discussion} \label{section:discussion}

\begin{figure}
\begin{center}
\includegraphics[width=10cm]{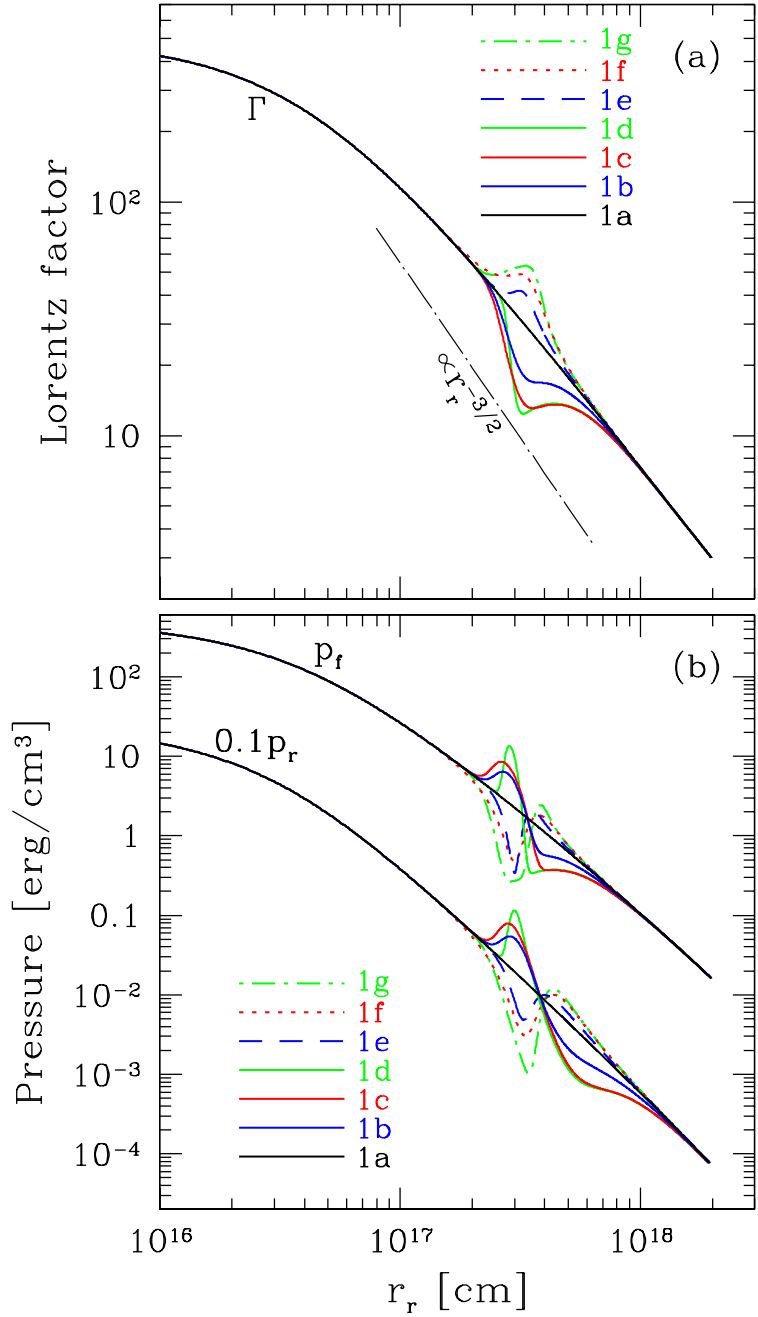}
\caption{
Shown together is the blast wave dynamics of examples 1a, 1b, 1c, 1d, 1e, 1f, and 1g. 
The notations are the same as in Figure~\ref{fig:alldyn_1a1b1c}. 
Panel (a) shows the $\Gamma$ curves, and the panel (b) shows the $p_f$ and $p_r$ curves. 
The $p_r$ curves are multiplied by a factor of 0.1, 
in order to avoid an overlap with the $p_f$ curves.
} 
\label{fig:dyn_1ato1g}
\end{center}      
\end{figure}

\begin{figure}
\begin{center}
\includegraphics[width=17cm]{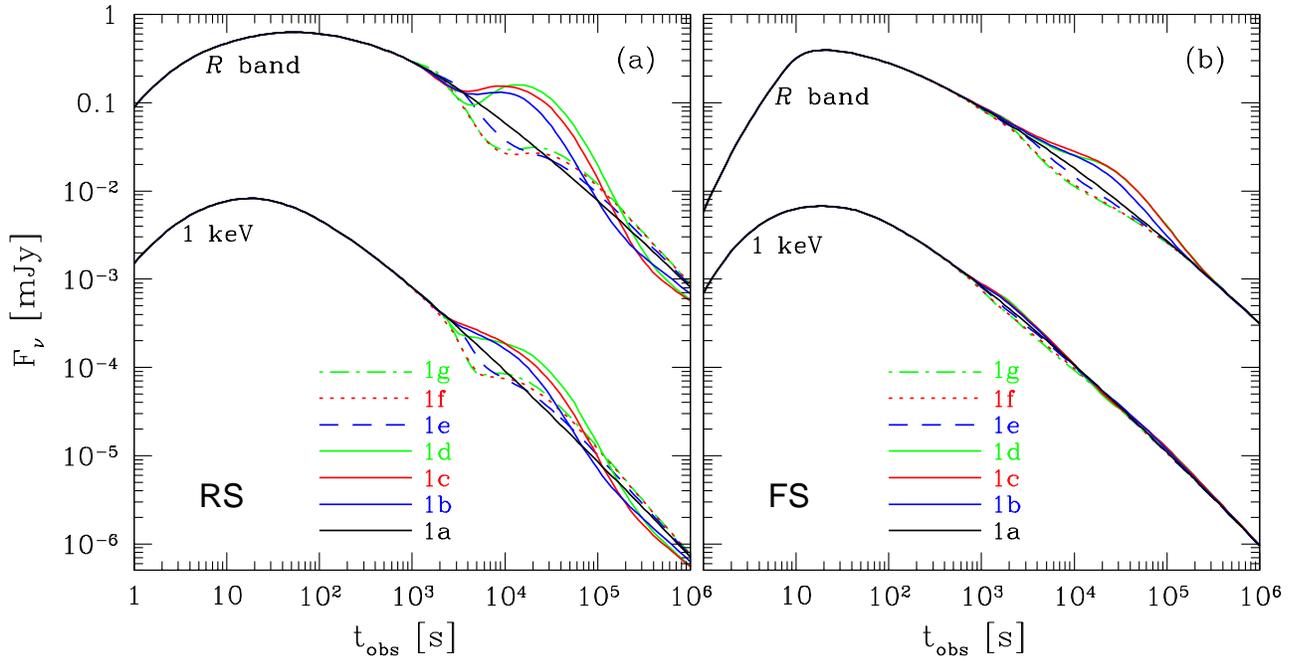}
\caption{
Same as in Figure~\ref{fig:cRX_1a1b1c}, 
but for examples 1a, 1b, 1c, 1d, 1e, 1f, and 1g. 
} 
\label{fig:cRX_1ato1g}
\end{center}      
\end{figure}

In this paper, we have investigated the blast wave dynamics and afterglow light curves 
of a GRB blast wave encountering various density structures in the surrounding ambient 
medium. We use the semi-analytic formulation of relativistic blast waves (presented in U11) 
to find an accurate numerical solution for both the FS and RS dynamics and employ the 
Lagrangian description of the blast waves (presented in U12) to perform a sophisticated 
calculation of afterglow emission. In particular, we consider numerical examples with 
density bumps, voids, or steps in the ambient medium and present their FS and RS 
dynamics and light curves. We form groups of examples in order to provide an efficient 
comparison among them and to help readers understand how the FS and RS dynamics should 
behave when the blast waves encounter those density structures. Giving a detailed explanation 
on their dynamics, we stress on the important role of adiabatic cooling of blast waves 
and then propose a number of consistency tests that correct modeling of blast waves 
needs to satisfy. We also present examples with a short-lived RS and show that the 
same consistency tests need to be satisfied for the FS dynamics. All these tests should serve 
as an accuracy indicator for any future (analytical, numerical, or hydrodynamical) modeling of 
blast waves with a long/short-lived RS. We point out that the RS dynamics responding 
to density structures in the ambient medium is studied extensively for the first time 
here. Also, the FS dynamics responding to density voids is presented here for the first time. 
As a summary, in Figure~\ref{fig:dyn_1ato1g}, we show together the blast wave dynamics of 
examples with a long-lived RS for bumps (i.e., 1b, 1c, and 1d) and voids (i.e., 1e, 1f, 
and 1g); the example 1a with our constant base density is also shown together. The FS and 
RS dynamics exhibit again an excellent agreement among examples beyond the bump/void region. 
In Figure~\ref{fig:cRX_1ato1g}, we show together the FS and RS light curves of these 7 examples. 
It appears that the RS light curves can produce some interesting observed features such as 
re-brightenings, dips, and slow wiggles, while the FS light curves in general produce 
weaker and smoother features, which may not be able to interpret significant bumps and wiggles
observed in some afterglows.

Why does the bump/void leave such a different imprint on the FS and RS light curves? 
The question is intriguing since the $p_f$ and $p_r$ curves exhibit a similar degree of 
response signature for bumps or voids (as one can see from Figure~\ref{fig:dyn_1ato1g}). 
We first recall that different microphysics parameters are adopted for the FS and RS; 
$\epsilon_e=10^{-1}$ and $\epsilon_B=10^{-2}$ for the RS light curves and 
$\epsilon_e=10^{-2}$ and $\epsilon_B=10^{-4}$ for the FS light curves, 
so that the FS and RS spectral density fluxes become comparable to each other. 
As a first step, we examine the effect of microphysics parameters on the response feature. 
In general, a higher value of $\epsilon_e$ or $\epsilon_B$ gives a weaker response signature 
on both the FS and the RS light curves. If we increase $\epsilon_e$, the Compton parameter $Y$ 
increases, which makes the electrons cool faster. If we increase $\epsilon_B$, the magnetic field 
strength $B$ increases, which also makes the electrons cool faster. In either case, the cooling 
Lorentz factor of our Lagrangian shells decreases faster. As a result, a smaller number of 
recently shocked shells have the electrons that can contribute to the X-ray or {\it R} band emission. 
In other words, for a higher $\epsilon_e$ or $\epsilon_B$ value, a smaller region of 
the blast wave responds to bumps or voids, 
leaving a weaker signature on the light curves. Conversely, a lower value of $\epsilon_e$ or 
$\epsilon_B$ gives a stronger response signature on both the FS and the RS light curves. 
For the same parameters $\epsilon_e=10^{-1}$ and $\epsilon_B=10^{-3}$ adopted for both shocked 
regions, the RS light curves show a stronger signature than in Figure~\ref{fig:cRX_1ato1g} 
whereas the FS light curves show an even weaker (nearly negligible) signature than 
in Figure~\ref{fig:cRX_1ato1g}. Therefore, the different microphysics parameters that 
we adopted for the FS and RS regions do not provide an explanation for the question. 
In fact, the microphysics parameters are chosen in such a way that the bump/void signature
becomes enhanced on the FS light curves as compared to the RS light curves.

The different signatures imprinted on the FS and RS light curves can be mainly understood 
by considering the influence of injection Lorentz factor of electrons. As the blast wave 
encounters density bumps, the $p_f$ curve rises, but the $\Gamma$ curve declines, which decreases 
the injection Lorentz factors in the FS fresh shells. Hence, there is competition between 
the rising energy density of shocked gas and the declining injection energy of electrons 
in the FS region. On the other hand, while encountering density bumps, the $p_r$ and $\gamma_{43}$ 
curves rise together. Thus, there is synergy between the rising energy density of shocked gas and 
the rising injection energy of electrons in the RS region. Similarly, as the blast wave 
encounters density voids, the FS region has competition between the decreasing energy density of 
shocked gas and the flattening (or increasing) injection energy of electrons. The RS region 
has again synergy between the declining energy density of shocked gas and the decreasing 
injection energy of electrons. This is the main source of the FS/RS asymmetry in response
to the bump/void structures in the ambient medium.

Another (secondary) reason for different imprints on the FS and RS comes from the intrinsic 
difference of the FS and RS propagation direction in the co-moving fluid frame. The FS 
propagates forward in the co-moving frame, so that the photons emitted from fresh shells 
are ``mixed'' together with the photons coming from old shells inside the shocked region. 
This photon mixing among the fresh and old shells tends to smooth out the bump/void signature 
in the FS light curves. On the other hand, the RS propagates backward in the co-moving frame, 
so that the photons emitted from fresh shells should arrive later than the photons coming from 
old shells inside the shocked region. Thus, there is no photon mixing among the fresh and old 
shells, which helps the RS light curves keep the bump/void signature.

The FS/RS asymmetry presented in this paper provides an important insight on the origin of afterglow 
emission. In reality, the observed light curves do not distinguish the contribution of the RS from the FS. 
In the standard framework of afterglow production by a blast wave with a short-lived RS, the contribution 
from the RS region after the RS crossing the end of the ejecta has a distinguishable temporal index 
($\sim -2$) in the observed light curves. However, when a blast wave with a long-lived RS is considered, 
the contribution from the RS region can not be easily 
distinguished from the FS contribution by simply looking at the temporal index of the observed light 
curves. Under these circumstances, one possible way to distinguish one from the other would be 
to investigate in detail their response features due to the inhomogenities in the ambient medium 
(as we do here in this paper) or in the ejecta stratification (as presented in U12). 
As we showed in this paper and U12, these inhomogenities generally leave a lot stronger signatures 
in the RS light curves than in the FS light curves. Therefore, these results suggest that the observed 
light curves with strong features favor toward the RS origin. One may think that the FS emission could 
account for some weak features. We believe that the spectral evolution of the FS and RS emission during 
the bump/void encounters may help further distinguish one from the other in the future, which will be 
investigated elsewhere. For the observed light curves without any noticeable feature, we believe that 
the future measurements of ``polarization degree curves'' for a longer time scale, say, up to hours or days, 
would hold a key to distinguish the contribution of the RS from the FS.

We also add another important remark regarding the possibility to differentiate between 
the two popular scenarios to account for the observed light curve features, 
i.e., the inhomogenities in the ambient medium (the current paper) or in the ejecta structure (U12). 
These two popular scenarios in fact give rise to different signatures. 
It appears that, for the RS emission, the inhomogenities in the ambient medium leave a stronger 
feature in the optical light curves than in the X-ray light curves, whereas the inhomogenities in 
the ejecta structure leave a stronger feature in the X-ray light curves than in the optical light curves. 
This tendency is shown in the current paper and U12, but requires a more detailed investigation 
to enable a better understanding on the underlying physics. If this is indeed the case, and if 
the observed emission is dominated by the RS emission,
then the multi-wavelength afterglow observations will be able to differentiate the two scenarios.

The bump/void structures presented in our models are essentially concentric shells of high or low 
density medium stacked upon each other. However, in reality, these structures should be three 
dimensional in nature, surrounded by an inter-clump medium (ICM). This 3-D aspect of the structures 
would then significantly alter the dynamics of the blast wave. It can be expected that the blast 
wave dynamics can be dominated by the uniform ICM rather than the bump/void structures if the clumps 
of these structures are sufficiently small. In this case, however, those small clumps would not leave 
significant features anyway, and the observed light curves would be dominated by the uniform ICM. 
For large clumps that could considerably affect the blast wave dynamics, we believe that our models 
can still reasonably apply when the clumps along the line of sight are as big as the visible area 
with an angular size $\sim 1/\Gamma$.

We also studied the step-like density structures in the paper. These structures may exist as 
a consequence of pre-ejection history from the central engine. When the relativistic blast wave 
encounters previously-ejected slowly moving denser shells, these shells could act as an ascending 
step-like structure. When the blast wave exits these denser shells, it experiences a descending 
step structure. In the case of a wind environment, a rapid change in the mass-loss rate 
in the past could create a step-like density structure. Note that these situations would probably 
create a relatively spherical density structure when viewed from the jetted blast wave. 
We stress that a descending step structure is proposed here to give a natural explanation 
for a dip feature -- a steepening followed by a flattening -- in GRB light curves.

%
%

\acknowledgments 
We thank the anonymous referee for many helpful comments and suggestions, which allowed us 
to improve the presentation of the paper. 
This work is supported by the China Postdoctoral Science Foundation through 
Grant No. 2013M540813, and National Basic Research Program (``973'' Program) of China 
under Grant No. 2014CB845800.

%
%


%
%

\end{document}